\documentclass{aastex}
\usepackage{emulateapj5}
\usepackage{graphics}
\newcommand{\bmath}[1]{\mbox{\boldmath\(#1\)}}
\renewcommand{\cite}[1]{\citeyear{#1}}
\bibpunct{\hspace{-4pt}}{}{}{a}{}{}

\submitted{{\rm Submitted to ApJL.}}

\begin{document}

\title{Pinpointing the massive black hole in the Galactic Center with gravitationally
lensed stars}

\author{Tal Alexander }

\affil{Space Telescope Science Institute, 3700 San Martin Drive, Baltimore, MD
21218}

\begin{abstract}
A new statistical method for pinpointing the massive black hole (BH) in
the Galactic Center on the IR grid is presented and applied to astrometric
IR observations of stars close to the BH. This is of interest for measuring
the IR emission from the BH, in order to constrain accretion models; for
solving the orbits of stars near the BH, in order to measure the BH mass
and to search for general relativistic effects; and for detecting the fluctuations
of the BH away from the dynamical center of the stellar cluster, in order
to study the stellar potential. The BH lies on the line connecting the
two images of any background source it gravitationally lenses, and so the
intersection of these lines fixes its position. A combined search for a
lensing signal and for the BH shows that the most likely point of intersection
coincides with the center of acceleration of stars orbiting the BH. This
statistical detection of lensing by the BH has a random probability of
\( \sim \! 0.01 \). It can be verified by deep IR stellar spectroscopy,
which will determine whether the most likely lensed image pair candidates
(listed here) have identical spectra. 
\end{abstract}

\keywords{Galaxy: center --- gravitational lensing --- infrared: stars}

\section{Introduction}

SgrA\( ^{\star } \), the unusual radio source in the Galactic Center (GC),
which was long suspected to be the central massive black hole (BH) of
the Galaxy, is now securely identified with the \( \sim \! 3\times 10^{6}\, M_{\odot } \)
dark compact mass that was detected dynamically by IR observations of stars
following a Keplerian velocity field (Eckart \& Genzel \cite{Eck97}; Ghez
et al. \cite{Ghe98}). The identification is based on the non-thermal spectrum
of SgrA\( ^{\star } \) (e.g. Serabyn et al. \cite{Ser97}), on its compactness
(Rogers et al. \cite{Rog94}), on its lack of detected motion (Backer \&
Sramek \cite{Bac99}; Reid et al. \cite{Rei99}) and on its position within
\( 0.1'' \) (\( 1''=0.04\, \mathrm{pc}\textrm{ } \) in the GC) of the
dynamical center of the Galaxy (Eckart \& Genzel \cite{Eck97}; Ghez et
al. \cite{Ghe98}). The nominal position of SgrA\( ^{\star } \) relative
to the stars was established by aligning the radio and IR grids to within
\( 0.03'' \) (\( 1\sigma  \) error) using 4 maser giants in the inner
\( 15'' \) of the GC (Menten et al. \cite{Men97}). Recent measurements
of the acceleration vectors of three stars very close to the BH provide
another dynamical method for localizing the BH and constrain the center
of acceleration to be inside an elongated \( \sim \! 0.03''\times 0.06'' \)
(\( \sim \! 1\sigma  \) error) region offset by \( 0.05'' \) from the
nominal position of SgrA\( ^{\star } \) (Ghez et al. \cite{Ghe00}). Neither
of these positions coincide with any observed IR point source. The extreme
under-luminosity of the BH across the spectrum poses a challenge to theoretical
accretion models (e.g. Narayan et al. \cite{Nar98}) and motivates the
continuing effort to detect, localize and measure the flux of the BH in
bands other than the radio. 

A precise determination of the BH position is also relevant for solving
the orbits of stars near the BH, which can be used to measure the BH mass
\( M_{\bullet } \) and the distance to the GC \( R_{0} \) (Jaroszynski
\cite{Jar99}; Salim \& Gould \cite{Sal99}) and to search for general
relativistic effects (Jaroszynski \cite{Jar98}). The BH will not lie exactly
at the dynamical center of the stellar cluster due to its Brownian motion.
A detection of this offset, which is expected to be of order few mas (Bahcall
\& Wolf \cite{Bah76}), can be used to estimate \( M_{\bullet } \) and
to probe the stellar potential and the formation history of the system.

Gravitational lensing of background stars by the BH offers another, independent
way for localizing the BH, which is free of assumptions about \( M_{\bullet } \),
\( R_{0} \) or the distribution of stars around the BH. The BH lies on
the line connecting the two images of any background source it gravitationally
lenses, and so the intersection of these lines fixes its position. The
most reliable way to identify lensed image pairs is by their spectra, which
should be identical up to differences caused by non-uniform extinction.
At present spectra are available for all stars brighter than \( \mathrm{K}=11 \)
mag, but only for a few stars fainter than \( \mathrm{K}\gtrsim 13 \)
mag. No spectral identification of lensing has been reported yet. However,
it is possible to use astrometric and photometric data to search statistically
for the signature of lensing. This is motivated by rough estimates that
suggest that there are several hundreds of distant background stars within
\( 2'' \) of SgrA\( ^{\star } \) (Alexander \& Loeb \cite{Ale01}).

A point mass lens is assumed here since the total stellar mass around the
BH on the scale of interest is very small compared to \( M_{\bullet } \).
The possibility of one of the images being amplified by a star near the
BH (Alexander \& Loeb \cite{Ale01}) is neglected.

\section{Method}

\label{s:method}

The two gravitationally lensed images of a background source lie on one
line with the point mass lens and the source, one on either side of the
lens. The angular distances of the images from the lens, \( \theta _{1,2} \),
and that of the source, \( \beta  \), are related by \( \beta =\theta _{1,2}-\theta ^{2}_{\mathrm{E}}/\theta _{1,2} \)
(\( \theta _{2}<0 \) by definition), where \( \theta _{\mathrm{E}} \)
is the Einstein angle, \( \theta ^{2}_{\mathrm{E}}=4GM_{\bullet }D_{ls}/D_{ol}D_{os}c^{2}=-\theta _{1}\theta _{2} \),
\( M_{\bullet } \) is the mass of the BH and \( D_{ls} \), \( D_{ol} \)
and \( D_{os} \) are the lens-source, observer-lens and
observer-source distances, respectively. The images are magnified by
\( A_{1,2}=\left| 1-\theta ^{4}_{\mathrm{E}}/\theta _{1,2}^{4}\right|
^{-1} \) (e.g. Schneider, Ehlers \& Falco \cite{Sch92}). It then
follows that the ratios of the angular distances, the projected radial
and transverse velocities and the fluxes of the two images obey the
relations \begin{equation}
\label{eq:ratios}
-\frac{\theta _{1}}{\theta _{2}}=-\frac{v_{r1}}{v_{r2}}=\frac{v_{t1}}{v_{t2}}=\sqrt{\frac{f_{1}}{f_{2}}}\, ,
\end{equation}
where the velocities are measured relative to the lens. These constraints
can be used to test whether two point sources are a pair of lensed images
using astrometric and photometric data. The challenge lies in detecting
a small number of true image pairs among the \( n \) observed stars against
the combinatorial background of the \( (n-1)n/2 \) possible pairs in the
presence of measurement errors. 

The constraints on the lensed image pair can be expressed more robustly
in terms of the projections \begin{equation}
\label{eq:C}
C_{\theta }=-\frac{\bmath {\theta _{1}}\cdot \bmath {\theta _{2}}}{\theta _{1}\theta _{2}}=1\qquad \mathrm{and}\qquad C_{w}=\frac{\bmath {\theta }\cdot \bmath {w}}{\theta w}=1\, ,
\end{equation}
 where \( \bmath {\theta }\equiv (\pm \theta _{1},\theta _{2}) \) and
\( \bmath {w} \) stands for the two-component vectors similarly constructed
from \( v_{r1,2} \), \( v_{t1,2} \) and \( f^{1/2}_{1,2} \), with the
signs chosen so that \( C_{w}=1 \) for a lensed image pair. The definition
of \( C_{w} \) utilizes the fact that stellar positions are the best measured
quantity in the data analyzed here. In the presence of measurement errors
Eq.~\ref{eq:C} is replaced by a score function,

\begin{equation}
\label{eq:S2}
S_{12}^{2}=\sum _{w=\theta ,v_{r},v_{t},\sqrt{f}}\left( \frac{C_{w}-1}{\Delta C_{w}}\right) ^{2}\, ,
\end{equation}
where \( \Delta C_{w} \) is the error in \( C_{w} \) due to the measurement
errors. The smaller the score, the more likely it is that the two stars
are a lensed image pair. 

It is not known how many lensed image pairs there are among the stars observed
near the BH, if any, and so the search for the position of the BH and for
the lensed image pairs has to be conducted simultaneously by looking for
deviations from a random (unlensed) stellar field. This proceeds by enumerating
over all possible locations for the BH around the nominal origin within
a specified search field, which is very small compared to the entire stellar
field; recalculating \( \theta  \), \( v_{t} \) and \( v_{r} \) relative
to the trial BH location; calculating \( S_{ij}^{2} \) for each of the
possible pairs \( (i,j) \); sorting the scores; calculating the likelihood
\( L \) for the best \( n/e \) scores under the assumption of a random
distribution; and finally, identifying the position of the BH as that with
the \emph{minimal} likelihood (ML) in random. The choice of the best \( n/e \)
scores minimizes the number of dependent pairs in random draws. Dependent
pairs, such as \( (i,j) \) and \( (i,k) \), increase the noise because
they cannot both be lensed image pairs. Formal confidence limits on the
ML position are calculated in the same way as for maximal likelihood parameter
estimation.

The probability distribution function (PDF) of the best \( n/e \) scores
and their likelihood in unlensed data are estimated directly by Monte-Carlo
(MC) simulations. The simulations consist of re-shuffling the actual data
by rotating each stellar position and velocity vector by random angles
and by randomly permuting the fluxes among the stars. In addition, the
entire data is randomly shifted within the search field to avoid biases
with respect to the nominal origin. This procedure conserves the averaged
radial density distribution, velocity field and luminosity function while
randomizing any angular correlations.

Some of the pairs are filtered out to further suppress the combinatorial
noise. A fraction of the stars have spectral type identification and line-of-sight
velocity measurements. Pairs whose stellar types (early or late type) are
discrepant are not compared, and so are stars whose line-of-sight velocities
\( v_{z}\pm \Delta v_{z} \) are discrepant by more than \( \sqrt{\Delta ^{2}v_{z1}+\Delta ^{2}v_{z2}} \).
In addition, only pairs with \( \theta _{\mathrm{E},\mathrm{min}}<\sqrt{-\theta _{1}\theta _{2}}<\theta _{\mathrm{E},\mathrm{max}} \)
are considered. This is necessary for avoiding pairs whose distance from
the center is too small relative to the astrometric errors to allow a reliable
test of co-linearity, or pairs whose Einstein angle clearly exceeds the
upper limit \( \theta _{E}=1.75''\pm 0.20'' \) for a source at infinity,
based on estimates of the BH mass, \( M_{\bullet }=(3.0\pm 0.5)\times 10^{6}\, M_{\odot } \)
(Genzel et al. \cite{Gen00}) and the distance to the GC, \( R_{0}=8.0\pm 0.5 \)
kpc (Reid \cite{Rei93}). In practice, the results do not depend strongly
on the exact filtering criteria as long as the search field is not too
large.

The random probability for finding a ML smaller than that found in the
data is \( P_{\mathrm{rand}}=P_{\mathrm{data}}(<\! \mathrm{ML})N \), where
\( P_{\mathrm{data}}(<\! \mathrm{ML}) \) is the probability of any \emph{given}
position in the search field to have a likelihood \( <\! \mathrm{ML} \),
which is calculated from the random PDF, and where \( N \) is the size
of the enumeration over the BH position. The actual value of \( N \) is
much smaller than the search field grid size (here \( 100^{2} \)) because
nearby locations are not independent due to the smoothing effect of the
measurement errors (Fig.~\ref{fig: shift}). However, it can be estimated
by \( N^{-1}\simeq \widehat{P}_{\mathrm{sim}}(<\! \mathrm{ML}) \), the
MC median of the likelihood probabilities \( P_{\mathrm{sim}}(<\! \mathrm{ML}) \)
in randomly re-shuffled fields constructed from the data. Inspection of
the likelihood plot in Fig.~\ref{fig: shift} suggests that there are
of order \( \mathrm{few}\times 10 \) resolution elements in the search
field, in agreement with the MC estimate \( N=28 \).

\section{Results}

\label{s:results}

The analysis is based on the compilation of Genzel et al. (\cite{Gen00}),
which includes all available astrometric observations of stars in the inner
\( 20'' \) of the GC. The compilation lists the stellar positions at a
common epoch, proper and radial velocities, when available (error weighted
averages if more than one measurement exists), K-band magnitudes and spectral
types, when known. The positions are the best measured quantity. Individual
errors are quoted only for the velocities. Genzel et al. (\cite{Gen00})
estimate a typical astrometric error of \( 10 \) mas (random and systematic).
Magnitudes are accurate to better than \( 0.1 \) mag for stars with \( \mathrm{K}<13 \)
mag, while fainter magnitudes are listed without errors to \( 0.1 \) mag
precision. A magnitude uncertainty of \( \Delta \mathrm{K}=0.1 \) mag
is assumed here for all stars. The sample is augmented by 20 stars observed
by Ghez et al. (\cite{Ghe98}) that were not included in the compilation
because of their uncertain velocity measurements, bringing the total to
295 stars. Of these, 116 stars have a complete set of measurements (position,
proper velocity and magnitude) and are used in the analysis.

\begin{figure*}[t]
{\centering \resizebox*{!}{0.29\textheight}{\includegraphics{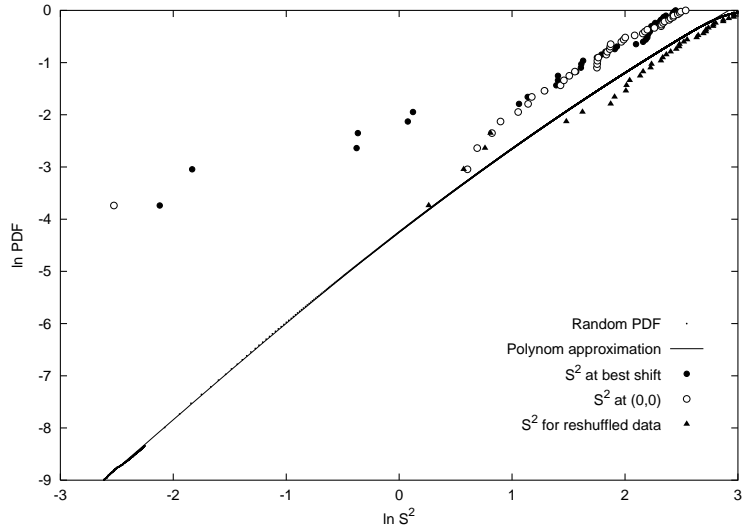}} \par}

\caption{\label{fig: pdf}A comparison of the low-\protect\( S^{2}\protect \)
tails of the probability distribution functions (PDFs) of the original
astrometric data, the best-fit shifted data, a random realization derived
from re-shuffling the data, and the random PDF derived from averaging \protect\( 10^{5}\protect \)
random realizations of re-shuffled data. The polynom used to approximate
the random PDF is also shown.}
\end{figure*}

\begin{figure*}[t]
{\centering \begin{tabular}{cc}
\resizebox*{!}{0.32\textheight}{\rotatebox{270}{\includegraphics{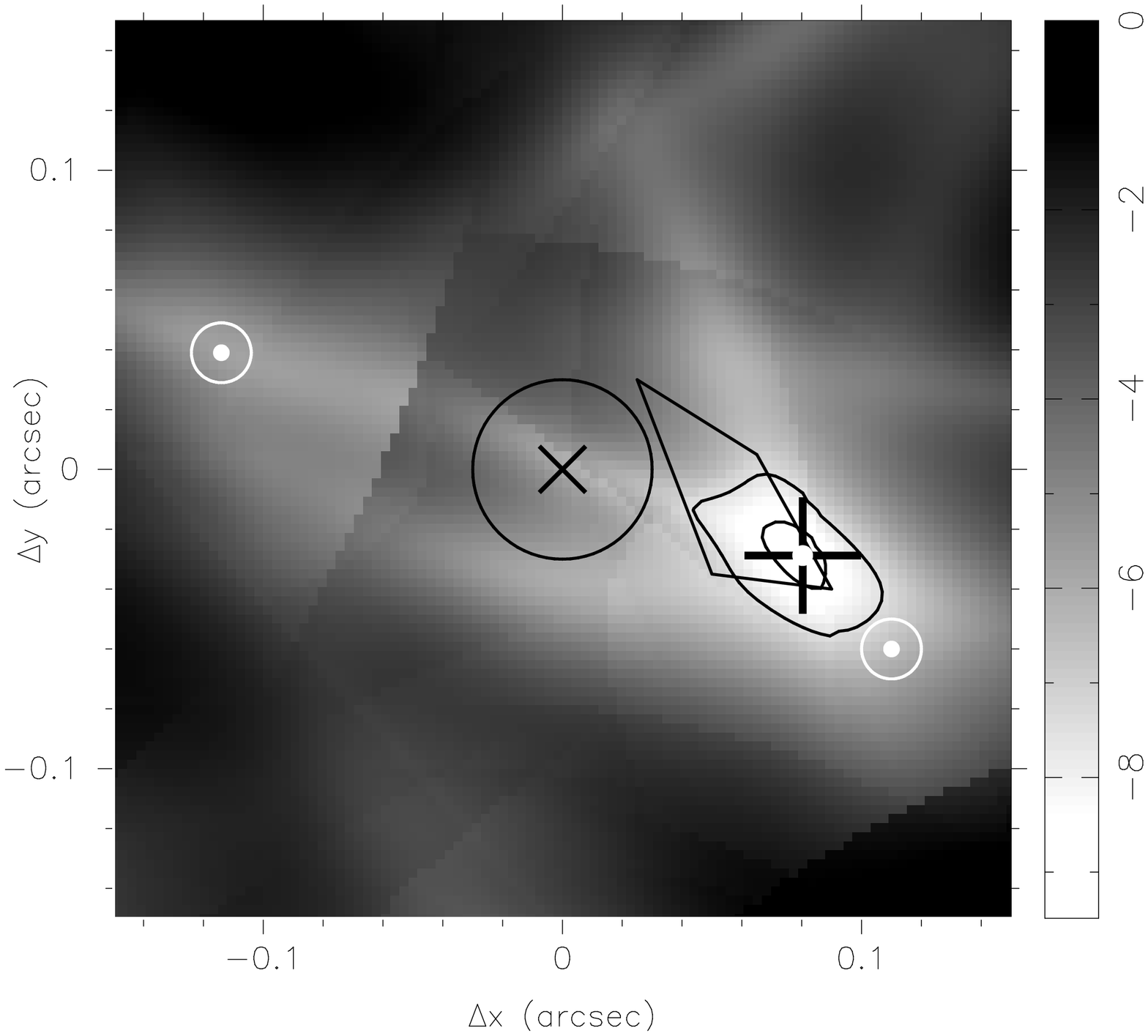}}} &
\resizebox*{!}{0.32\textheight}{\rotatebox{270}{\includegraphics{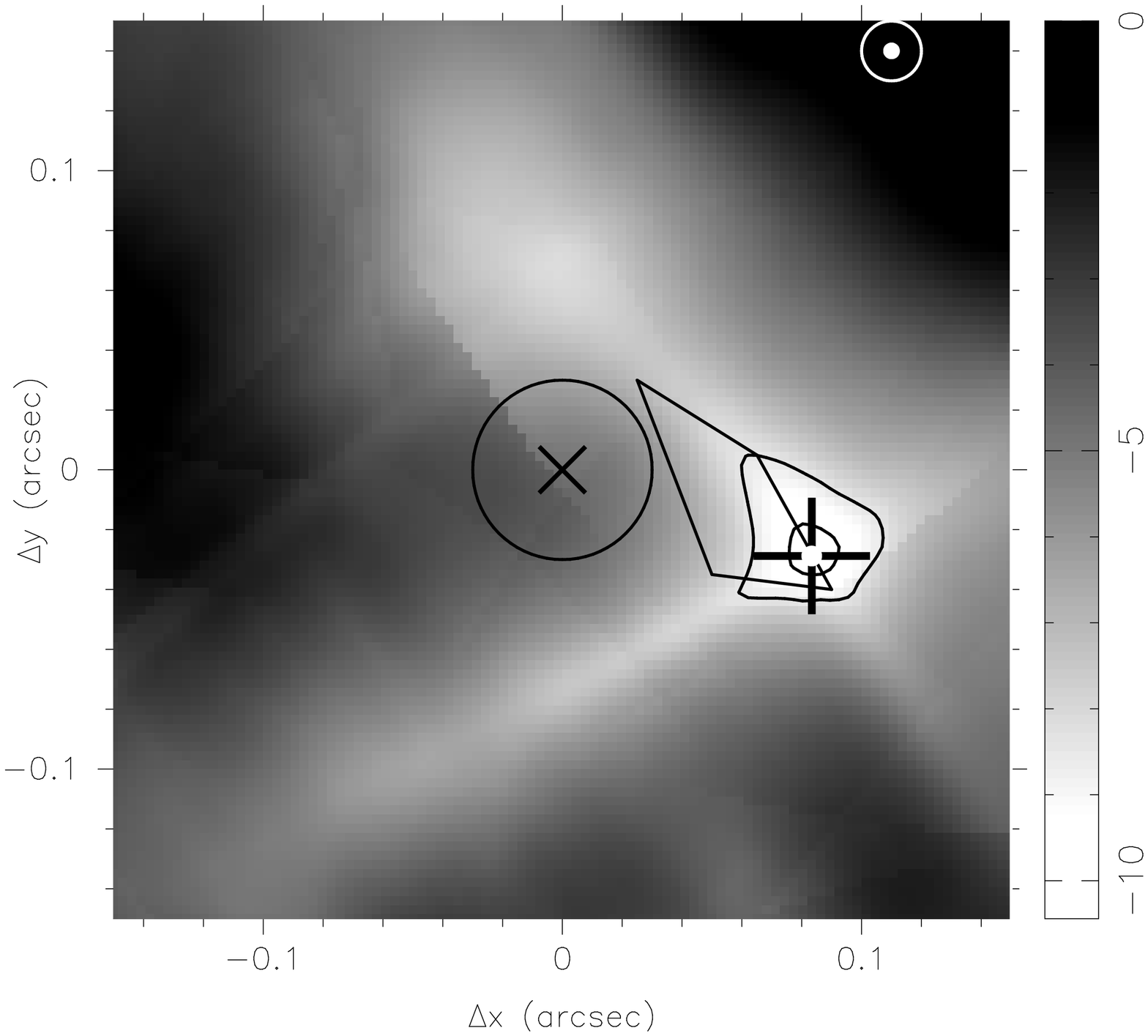}}} \\
\end{tabular}\par}

\caption{\label{fig: shift}Left: A gray scale plot of \protect\( \ln \mathrm{ML}\protect \)
(shifted to 0 at the maximum) for 116 stars, as function of the shift in
the astrometric grid \protect\( (\Delta x,\Delta y)=(\Delta \mathrm{RA},\Delta \mathrm{Dec})\protect \)
over the central \protect\( 0.3''\times 0.3''\protect \) search field.
The X in the center is the origin according to the IR/radio alignment with
its \protect\( 1\sigma \protect \) error circle (Menten et al. \cite{Men97}).
The polygon is the \protect\( \sim \! 1\sigma \protect \) error region
for the center of acceleration (Ghez et al. \cite{Ghe00}). The dots are
the observed IR sources with their 10 mas errors circles (Genzel et al.
\cite{Gen00}). The most likely shift at \protect\( (+0.076'',-0.027'')\protect \)
is indicated by a cross with \protect\( 1\sigma \protect \) and \protect\( 2\sigma \protect \)
confidence level contours. Right: As in left panel, for a mock field of
110 unlensed stars and 3 lensed image pairs, with the BH at \protect\( (+0.076'',-0.027'')\protect \).}
\end{figure*}

The results presented here are based on \( 10^{5} \) MC runs with filtering
criteria \( \theta _{\mathrm{E},\mathrm{max}}=2.1'' \) (a \( 2\sigma  \)
margin above the maximal value of \( \theta _{\mathrm{E}} \)) and \( \theta _{\mathrm{E},\mathrm{min}}=1.0'' \).
The latter corresponds to a chance co-linearity probability of \( \sim \! 0.3 \)
(hence the importance of using the additional constraints on the velocities
and fluxes). Figure~\ref{fig: pdf} compares the random PDF with the approximating
polynom that is used for calculating the likelihood, with the distribution
of the best \( n/e \) scores in one of the randomly re-shuffled realizations,
with the scores at the nominal origin and with those at the most likely
BH location. Figure \ref{fig: shift} shows the likelihood across the
\( 0.3''\times 0.3'' \) search field. The position of the ML at \( \Delta \mathrm{RA}=+0.076'' \),
\( \Delta \mathrm{Dec}=-0.027'' \) relative to the nominal origin lies
within the error region on the center of acceleration (Ghez et al. \cite{Ghe00})
and is \( 2.7\sigma  \) away from the nominal origin. The ML has a random
probability of \( P_{\mathrm{rand}}\sim \! 0.01 \). The position of the
ML is insensitive to moderate changes in \( \theta _{\mathrm{E},\mathrm{min}} \),
\( \theta _{\mathrm{E},\mathrm{max}} \), or to whether the three innermost
stars are included in the data set or not, or to higher assumed values
of \( \Delta \mathrm{K} \) up to \( 0.5 \) mag, which may better describe
the errors in the flux ratio given the patchy nature
of the extinction in the inner GC (Blum, Sellgren \& DePoy \cite{Blu96}).
However, \( P_{\mathrm{rand}} \) increases when larger errors are assumed.
The results are sensitive to the values adopted for the velocities. 90
stars of the compiled data were observed with the Keck telescope at high
resolution (\( \gtrsim \! 2 \) mas) over a baseline of 2 yr (Ghez et al.
\cite{Ghe98}). The ML analysis of the original Keck data has a global
minimum at the edge of the search field and only a weak local minimum near
\( (+0.08'',-0.03'') \). The difference is due to the compiled velocities,
which are the average of the high resolution, short baseline astrometry
and of lower-resolution, longer baseline astrometry.

Table \ref{tbl:pairs} lists the ten most-likely image pair candidates.
At present none of them have spectra for both images. The results of simulations
with mock stellar fields with properties similar to those of the data (Fig.
\ref{fig: shift}) confirm that the ML analysis has a high rate of success
in finding the position of the BH with as few as 3 image pairs in the data.
The simulated image pairs are usually ranked very close to the top of the
list, but some highly ranked candidates are false identifications.

\begin{table*}[t]

\caption{\label{tbl:pairs}The ten most likely image pair candidates.}

\centering

\begin{tabular}{cccccllrrrrll}
 {\small \( S_{12}^{2} \)}&
 {\small \( \theta _{\mathrm{E}} \)\( ^{a} \)}&
 {\small \( A_{1} \)\( ^{b} \)}&
 {\small \( \theta _{1} \)\( ^{a} \)}&
 {\small \( \theta _{2} \)\( ^{a} \)}&
 {\small \( K_{1} \)\( ^{c} \)}&
 {\small \( K_{2} \)\( ^{c} \)}&
 {\small \( v_{t1} \)\( ^{d} \)}&
 {\small \( v_{t2} \)\( ^{d} \)}&
 {\small \( v_{r1} \)\( ^{d} \)}&
 {\small \( v_{r2} \)\( ^{d} \)}&
 {\small \#1\( ^{e} \)}&
 {\small \#2\( ^{e} \)}\\
\hline
{\small 0.12}&
 {\small 1.09}&
 {\small 1.05}&
 {\small 2.29}&
 {\small 0.51}&
 {\small 11.6\( ^{\dag } \)}&
 {\small 14.8} &
 {\small -253}&
 {\small -118}&
 {\small +127}&
 {\small +53}&
 {\small 58}&
 {\small 13}\\
 {\small 0.14}&
 {\small 1.61}&
 {\small 3.82}&
 {\small 1.73}&
 {\small 1.49}&
 {\small 13.4\( ^{\dag } \)}&
 {\small 13.9 }&
 {\small -118}&
 {\small -129}&
 {\small -94}&
 {\small +11}&
 {\small 44}&
 {\small 34}\\
 {\small 0.71}&
 {\small 1.23}&
 {\small 1.03}&
 {\small 2.94}&
 {\small 0.51}&
 {\small 11.4\( ^{\dag } \)}&
 {\small 14.8} &
 {\small -176}&
 {\small -118}&
 {\small -73}&
 {\small +53}&
 {\small 70}&
 {\small 13}\\
 {\small 0.72}&
 {\small 2.00}&
 {\small 2.29}&
 {\small 2.31}&
 {\small 1.73}&
 {\small 12.5\( ^{\dag } \)}&
 {\small 13.3 }&
 {\small -314}&
 {\small -191}&
 {\small -152}&
 {\small -57}&
 {\small 64}&
 {\small 39}\\
 {\small 1.11}&
 {\small 1.29}&
 {\small 1.04}&
 {\small 2.98}&
 {\small 0.56}&
 {\small 11.9 }&
 {\small 15.0 }&
 {\small +121}&
 {\small +71}&
 {\small +131}&
 {\small -227}&
 {\small 72}&
 {\small 12}\\
 {\small 1.14}&
 {\small 1.09}&
 {\small 2.15}&
 {\small 1.28}&
 {\small 0.94}&
 {\small 12.5 }&
 {\small 12.9 }&
 {\small +345}&
 {\small -49}&
 {\small -39}&
 {\small +202}&
 {\small 28}&
 {\small 20}\\
 {\small 2.82}&
 {\small 1.73}&
 {\small 2.21}&
 {\small 2.01}&
 {\small 1.49}&
 {\small 12.9 }&
 {\small 13.9 }&
 {\small -75}&
 {\small -129}&
 {\small +134}&
 {\small +11}&
 {\small 49}&
 {\small 34}\\
 {\small 3.18}&
 {\small 1.73}&
 {\small 1.03}&
 {\small 4.25}&
 {\small 0.71}&
 {\small 11.2 }&
 {\small 14.0} &
 {\small +146}&
 {\small +104}&
 {\small -24}&
 {\small +104}&
 {\small 93}&
 {\small 15}\\
 {\small 3.57}&
 {\small 1.07}&
 {\small 1.03}&
 {\small 2.61}&
 {\small 0.44}&
 {\small 12.4\( ^{\dag } \)}&
 {\small 15.0 }&
 {\small +444}&
 {\small -248}&
 {\small -335}&
 {\small +384}&
 {\small 67}&
{\small  \phantom{1}9}\\
 {\small 3.59}&
 {\small 1.21}&
 {\small 1.05}&
 {\small 2.61}&
 {\small 0.56}&
 {\small 12.4\( ^{\dag } \)}&
 {\small 15.0} &
 {\small +444}&
 {\small +71}&
 {\small -335}&
 {\small -227}&
 {\small 67}&
 {\small 12}\\
\hline
\multicolumn{13}{l}{{\footnotesize \( ^{a} \)Einstein angle and radial distances in arcseconds,
relative to best fit BH position.} }\\
\multicolumn{13}{l}{{\footnotesize \( ^{b} \)The amplification of the brighter image. For
the second image \( A_{2}=A_{1}-1 \).}}\\
\multicolumn{13}{l}{{\footnotesize \( ^{c} \)K-band magnitudes. A spectrum also exists if
marked by (\( \dag  \)).}}\\
\multicolumn{13}{l}{{\footnotesize \( ^{d} \)Radial and transverse velocities in \( \mathrm{km}\, \mathrm{s}^{-1} \),
relative to best fit BH position.}}\\
\multicolumn{13}{l}{{\footnotesize \( ^{e} \)Row number of source in table 1 of Genzel et
al. (\cite{Gen00}).} }\\
\hline
\end{tabular}
\end{table*}

\section{Discussion and summary}

\label{s:discuss}

The results of the combined search for a lensing signal and for the BH
are intriguing because the most likely BH position coincides with the
center of acceleration and because of the low random probability of
the lensing signal. However, spectroscopic confirmation of the lensed
image pair candidates is essential since the statistical significance
of this result is sensitive to uncertainties in the measurement errors
and because the location of the ML depends on the values adopted for
the stellar velocities. More generally, a comprehensive spectroscopic
search for sources with matching spectra is the best method for
detecting lensing since it does not assume any specific mass
distribution for the lens. Many of the lensed image pair candidates
listed in Table~\ref{tbl:pairs} have small amplifications, \(
A_{2}=A_{1}-1\ll 1 \), with a 2 mag average difference between the two
images, and so none of the second images have spectral
identifications. Until deeper stellar spectroscopy is available, the
statistical analysis could be improved if individual errors were made
available for all the measured quantities. It should be noted that the
4 maser giants used for aligning the radio and IR grids are themselves
not stationary (the closest to the center, IRS7, has a proper motion
of \( 4\, \mathrm{mas}\, \mathrm{yr}^{-1} \), Genzel et
al. \cite{Gen00}).  Uncorrected, this could lead to small systematic
shifts between astrometric observations taken at different
epochs. Such shifts are unimportant for determining the dynamical
center, but could introduce errors in astrometric lensing searches in
heterogeneous compilations, especially for very high resolution data.

To summarize, astrometric and photometric observations of stars very near
the BH in the GC can be used to pinpoint the BH on the IR grid with as
few as 3 lensed image pairs in the field to an accuracy comparable with
that of dynamical methods. An analysis of all available astrometric data
indicates that the most likely position of the BH coincides with the center
of acceleration of stars orbiting the BH. Lensing by the BH has been detected
statistically with a random probability of \( \sim \! 0.01 \). This result
can be verified by deep stellar spectroscopy.

\acknowledgements{I am grateful to M. Fukugita, A. Loeb, P. Sackett and E. Quataert for
helpful discussions.}

\end{document}